\documentclass[superscriptaddress,pra]{revtex4}
\def\<{\langle }
\def\>{\rangle }

\def\Tr{\hbox{Tr} }
\def\map#1{\mathscr{#1}}
\def\sH{\mathcal H}
\usepackage{amsmath,mathrsfs}
\usepackage{amssymb}
\usepackage{amsthm}
\usepackage{graphicx}
\usepackage{epsfig}
\usepackage{subfigure}
\usepackage{hyperref}
\begin{document}

\title{Superbroadcasting of harmonic oscillators mixed states}
% Force line breaks with \\

\author{Giacomo M. D'Ariano} 
\email{dariano@unipv.it}
\author{Paolo Perinotti}
\email{perinotti@fisicavolta.unipv.it}
\affiliation{Dipartimento di Fisica ``A. Volta'' and CNISM, via Bassi
  6, I-27100 Pavia, Italy.}  
\author{Massimiliano F. Sacchi}
\email{msacchi@unipv.it}
\affiliation{Dipartimento di Fisica ``A. Volta'' and CNISM, via Bassi
  6, I-27100 Pavia, Italy.}  \affiliation{CNR - Istituto Nazionale per
  la Fisica della Materia, Unit\`a di Pavia, Italy.}

\date{\today}% It is always \today, today,
             %  but any date may be explicitly specified

\begin{abstract}
  We consider the problem of broadcasting quantum information encoded
  in the displacement parameter for an harmonic oscillator, from $N$
  to $M>N$ copies of a thermal state. We show the Weyl-Heisenberg
  covariant broadcasting map that optimally reduces the thermal photon
  number, and we prove that it minimizes the noise in conjugate
  quadratures at the output for general input states. We find that
  from two input copies broadcasting is feasible, with the possibility
  of simultaneous purification ({\em superbroadcasting}).  
\end{abstract}

\maketitle

\section{Introduction}

The no-cloning theorem \cite{noclon} states that one cannot produce a
number of independent physical systems prepared in identical states
out of a smaller amount of systems prepared in the same state. This
fact challenged the scientific community to find transformations which
can approximate the cloning transformation with the highest possible
fidelity. A whole branch in the literature is devoted to this problem,
and optimal cloners have been found, for qubits
\cite{buzhill,gismass,bruss}, for general finite-dimensional systems
\cite{werner}, for restricted sets of input states
\cite{darlop,darmacc}, and for infinite-dimensional systems such as
harmonic oscillators---the so called continuous variables cloners
\cite{cerf}. In all these cases pure states have been considered. On
the other hand, the problem of cloning mixed states is somehow more
interesting, since a more general type of cloning transformation can
be considered---the so-called {\em broadcasting}---in which the output
copies are in a globally correlated state whose local ``reduced''
states are identical to the input states. This possibility has been
considered in Ref.  \cite{nobro}, where it has been shown that
broadcasting a single copy from a noncommuting set of density matrices
is always impossible. The no broadcasting theorem was the only
interesting result for mixed states for many years, and this led to
the belief that a generalization of the no-cloning theorem to mixed
states should hold. However, more recently, for qubits an effect
called {\em superbroadcasting} \cite{prl} has been discovered, which
consists in the possibility of broadcasting the state while even
increasing the purity of the local state, for at least $N\ge 4$ input
copies, and for sufficiently short input Bloch vector (and even for
$N=3$ input copies for phase-covariant broadcasting instead of
universal covariance \cite{pra}), and these superbroadcasting maps can
be mixed with depolarizing maps in order to get exact broadcasting.

In the present paper, we analyze the broadcasting of mixed states of
an harmonic oscillator by a signal-preserving map. More precisely,
this means that we consider a set of input states obtained by
displacing a fixed mixed state by a complex amplitude in the harmonic
oscillator phase space, while the broadcasting map is covariant with
respect to the (Weyl-Heisenberg) group of complex displacements. We
will focus first on displaced thermal states (which are equivalent to
coherent states that have suffered Gaussian noise), and then we will
show that all results hold for any covariant set of mixed states in
terms of noise of conjugated quadratures.

As we will see, superbroadcasting is possible for harmonic oscillator
mixed states \cite{njp,epl}, namely one can produce a larger number of
copies, which are locally purified on each output oscillator, and with
the same signal as the input. For displaced thermal states, for
example, superbroadcasting can be achieved for at least $N=2$ input
copies, with thermal photon number $\overline{n}_{in}\ge\frac13$,
whereas, for sufficiently large $\overline{n}_{in}$ at the input, one
can broadcast to an unbounded number $M$ of output copies. For
purification (i.e.  $M\le N$), quite surprisingly the purification
rate is $\overline{n}_{out}/\overline{n}_{in}=N^{-1}$, independently
on $M$.  The particular case of 2 to 1 for noisy coherent states has
been reported in Ref. \cite{ula}. We will prove also that perfect
broadcasting is possible, provided that one knows the input thermal
photon number.

\section{Covariant broadcasting for the Weyl-Heisenberg group}
\label{covar}
We consider the problem of broadcasting $N$ input copies of displaced
(generally) mixed states of harmonic oscillators (with annihilation
operators denoted by $a_0,\ a_1,...,a_{N-1}$) to $M$ output copies
(with annihilation operators $b_0,\ b_1,...,b_{M-1}$). In order to
preserve the signal, the broadcasting map $\map B$ must be covariant,
i.~e. in formula
\begin{equation}
  \map B(D(\alpha)^{\otimes N}\Xi D(\alpha)^{\dag\otimes N})=D(\alpha)^{\otimes M}\map B(\Xi)D(\alpha)^{\dag\otimes M},
\label{cov}
\end{equation}
where $D_c(\alpha)=\exp(\alpha c^\dag-\alpha^* c)$ denotes the
displacement operator, and $\Xi$ represents an arbitrary $N$-partite
state. It is useful to consider the Choi-Jamio\l kowski bijective
correspondence of completely positive (CP) maps $\map B$ from
$\sH_\mathrm{in}$ to $\sH_\mathrm{out}$ and positive operators $
R_{\map B}$ acting on $\sH_\mathrm{out}\otimes\sH_\mathrm{in}$, which
is given by the following expressions
\begin{equation}
\begin{split}
  &R_{\map B}=\map B\otimes \map I(|\Omega\>\<\Omega|)\;,\\
  &\map B(\rho)=\Tr_\mathrm{in}[(I_\mathrm{out}\otimes\rho^\tau )R_{\map
    B}]\;,
\end{split}
\end{equation}
where $|\Omega\>=\sum_{n=0}^{\infty}|\psi_n\>|\psi_n\>$ is a maximally
entangled vector of $\sH_\mathrm{in} ^{\otimes 2}$, and $X^\tau $ denotes
transposition of $X$ in the basis $|\psi_n\>$. In terms of the
operator $R_{\map B}$ the covariance property \eqref{cov} can be
written as
\begin{eqnarray} 
[R_{\map B},D(\alpha )^{\otimes M}\otimes D(\alpha
  ^*)^{\otimes N}]=0\;,\qquad \forall \alpha \in {\mathbb
    C}\;.\label{com}
\end{eqnarray}
In order to deal with this constraint we introduce the multisplitter
operators $U_a$ and $U_b$, that perform the unitary transformations
\begin{eqnarray}
  && U_a a_k U^\dag _a =\frac 1{\sqrt N}\sum_{l=0}^{N-1}e^{\frac{2\pi i kl}N} a_{l} \;,
  \nonumber \\& & 
  U_b b_k U^\dag _b =\frac 1{\sqrt M} \sum_{l=0}^{M-1}e^{\frac{2\pi i
      kl}M}b_l
  \;.\label{multisplit}
\end{eqnarray}
Notice that such transformations perform a Fourier transform over all
input and output oscillators.   
Moreover, we will make use of the squeezing transformation
$S_{a_0b_0}$ defined as follows
\begin{equation}
\begin{split}
  & [S_{a_0 b_0},a_n]=[S_{a_0 b_0},b_n]=0,\; n>0\\
  & S_{a_0 b_0}a_0^\dag S_{a_0b_0}^\dag =\mu a_0^\dag -\nu b_0 \;,\\&
  S_{a_0b_0}b_0S_{a_0b_0}^\dag =\mu b_0 -\nu a_0^\dag\;,
\end{split}
\label{ampli}
\end{equation}
with $\mu = \sqrt{M/(M-N)}$ and $\nu =\sqrt{N/(M-N)}$. The squeezing
transformation here acts as an hyperbolic transformation for just
oscillators $a_0$ and $b_0$, by leaving all other oscillators
unaffected. In terms of such operators, condition (\ref{com}) becomes
\begin{eqnarray}
[S^\dag _{a_0b_0}(U^\dag_b\otimes U^\dag _a) R_{\map B} (U_b\otimes U_a) S_{a_0b_0},
  D_{b_0}(\sqrt{M-N} \alpha )]=0\;.\label{commrel}
\end{eqnarray}
Hence, upon introducing an operator $B$ on oscillators
$b_1,...,b_{M-1},a_0,...,a_{N-1}$, the operator $R_{\map B}$ can be
written in the form
\begin{eqnarray}
R_{\map B}= (U_b\otimes U_a) S_{a_0b_0} (I_{b_0}\otimes B)
S^\dag _{a_0b_0}(U^\dag_b\otimes U^\dag _a).
\end{eqnarray}
Notice that $R_{\map B}\geq 0$ is equivalent to $B\geq 0$. The further
condition that $\map B$ is trace-preserving in terms of $R_{\map B}$
becomes $\Tr_{b}[R_{\map B}]=I_{a}$, $b$ and $a$ collectively denoting
all output and input oscillators, respectively. From the trace and
completeness relations for the set of displacement operators,
namely $\int d^2 \alpha \,D(\alpha )A D^\dag (\alpha )=\hbox{Tr}[A]I$,
and $A=\int d^2 \alpha \,\Tr[D^\dag (\alpha )A]D(\alpha )$, (see,
e.g., Ref. \cite{bm}), the condition $\Tr_{b}[R_{\map B}]=I_{a}$ is verified
iff
\begin{eqnarray}
\left(\prod _{i=0}^{M-1} \int d^2 \beta _i \right ) 
\left(\bigotimes _{i=0}^{M-1} D_{b_i}(\beta _i ) \right)S_{a_0b_0}(I_{b_0} \otimes B)S^\dag _{a_0b_0}
\left(\bigotimes _{i=0}^{M-1} D^\dag _{b_i}(\beta _i)\right)= I
\;.\label{tracep}
\end{eqnarray}
From the relation $D_{b_0}(\beta _0)S_{a_0b_0}=S_{a_0b_0}D_{b_0}(\mu
\beta _0)\otimes D_{a_0}^\dag (\nu \beta _0)$, one obtains the
condition
\begin{eqnarray}
  \hbox{Tr}_{b/b_0,a_0}[B]=\nu ^2
  I_{a/a_0}\;,
\label{trb}
\end{eqnarray}
where $a/a_i$ denote all the input oscillators apart from $a_i$, and
similarly for $b/b_i$.

We will now consider the map corresponding to
\begin{equation}
  B=\nu ^2 \,
  |0\>\<0|_{b/b_0}\otimes|0\>\<0|_{a_0}\otimes
  I_{a/a_0}\;.
\label{bcho}
\end{equation}
Applying the corresponding map $\map B$ to a generic $N$-partite state
$\Xi$ we get
\begin{equation}\label{optmap}
  \map B(\Xi)=\Tr_{a}[(I_b\otimes\Xi^\tau )(U_b\otimes U_a) S_{a_0b_0} (I_{b_0}\otimes B)
  S^\dag _{a_0b_0}(U^\dag _b\otimes U^\dag_a)]\,,
\end{equation}
which is equivalent to
\begin{equation}
  \map B(\Xi)=\Tr_{a}[(I_b\otimes U^\dag _a\Xi^\tau U_a) (U_b\otimes I_a) S_{a_0b_0} (I_{b_0}\otimes B)
  S^\dag _{a_0b_0}(U^\dag_b\otimes I_a)]\,.
\end{equation}
Using the expression in Eq.~\eqref{bcho} we obtain
\begin{equation}
  \map B(\Xi)=U_b\left\{\Tr_{a_0}[(I_{b_0}\otimes\xi^\tau _{a_0}) S_{a_0b_0} (I_{b_0}\otimes |0\>\<0|_{a_0})
    S^\dag _{a_0b_0}]\otimes|0\>\<0|_{b/b_0}\right\}U^\dag_b\,,\label{13}
\end{equation}
where $\xi^\tau =\Tr_{a/a_0}[U^\dag _a\Xi^\tau U_a]$. Notice that
\begin{equation}
\begin{split}
  \xi=&\int\frac{d^2\gamma}\pi
  D(\gamma)^\tau \Tr[(D_{a_0}(\gamma)^\dag\otimes I_{a/a_0})U_a^\dag\Xi^\tau 
  U_a]\\=&\int\frac{d^2\gamma}\pi
  D(\gamma)^\tau \Tr[U^*_a(D_{a_0}(\gamma)^*\otimes I_{a/a_0})U_a^\tau \Xi],
\end{split}
\end{equation}
and using the complex conjugate of Eq.~\eqref{multisplit} we have
\begin{equation}
\begin{split}
  \xi=&\int\frac{d^2\gamma}\pi D(\gamma)^\tau \Tr[D(\gamma^*/\sqrt{N})^{\otimes N} \Xi ]\\
  =&\int\frac{d^2\gamma}\pi D(\gamma)^\tau \Tr[(D_{a_0}(\gamma)^*\otimes
  I_{a/a_0}) U_a^\dag\Xi U_a]=\Tr_{a/a_0}[U_a^\dag\Xi U_a].
\end{split}
\end{equation}
Now, we can easily evaluate $S_{a_0b_0} (I_{b_0}\otimes
|0\>\<0|_{a_0}) S^\dag _{a_0b_0}$, by expanding the vacuum state as
\begin{equation}
  |0\>\<0|_{a_0}=\int\frac{d^2\gamma}\pi 
  e^{-\frac{|\gamma|^2}2}D_{a_0}(\gamma)\,,
\end{equation}
obtaining
\begin{equation}
  S_{a_0b_0} (I_{b_0}\otimes |0\>\<0|_{a_0}) S^\dag _{a_0b_0}=\int\frac{\nu^2d^2\gamma}{\pi} e^{-\frac{|\gamma|^2}2}D_{b_0}(\nu\gamma^*)\otimes D_{a_0}(\mu\gamma)\,.
\end{equation}
Hence, Eq. (\ref{13}) can be rewritten as
\begin{equation}
  \map B(\Xi)=\int\frac{d^2\gamma}\pi U_b (D_{b_0}
  (\gamma^*)\otimes|0\>\<0|_{b/b_0})U_b^\dag e^{-\frac{|\gamma|^2}{2\nu^2}}\Tr[D_{a_0}(\mu\gamma/\nu)\xi^\tau ]\,.
\end{equation}
As an example, we will now consider $N$ displaced thermal states
\begin{equation}
  \rho_\alpha\doteq\frac1{\bar n+1}D(\alpha)\left(\frac{\bar n}{\bar n+1}\right)^{a^\dag a}D(\alpha)^\dag\,,
\end{equation}
from which we want to obtain $M$ states, the purest as possible.
Thanks to the covariance property, it is sufficient to focus attention
on the output of $\rho_0^{\otimes N}$. For a tensor product of thermal
input states $\Xi=\rho_0^{\otimes N}$, exploiting the fact that
$U_a^\dag (\sum_{j=0}^{N-1}a^\dag_j a_j) U_a=\sum_{j=0}^{N-1}a^\dag_j
a_j$, we have
\begin{equation}
  \xi=\xi^\tau =\rho_0\,,
\end{equation}
and recalling the following expression for the thermal states
\begin{equation}
  \frac1{\bar n+1}\left(\frac{\bar n}{\bar n+1}\right)^{a^\dag a}=\int\frac{d^2\beta}\pi e^{-\frac{|\beta|^2}2(2\bar n+1)}D(\beta)\,,
\label{therm}
\end{equation}
we obtain 
\begin{equation}
\begin{split}
  \map B\left(\rho_0^{\otimes N}\right)=&\int\frac{d^2\gamma}\pi U_b
  (D_{b_0} (-\gamma^*)\otimes|0\>\<0|_{b/b_0})U_b^\dag
  e^{-\frac{|\gamma|^2}{2\nu^2}[\mu^2(2\bar n+1)+1]}\\
  &=\int\frac{d^2\gamma}{\pi\bar n'} U_b
  (|\gamma\>\<\gamma|_{b_0}\otimes|0\>\<0|_{b/b_0})U_b^\dag
  e^{-\frac{|\gamma|^2}{\bar n'}} \\
  &=\int\frac{d^2\gamma}{\pi\bar n'}
  |\gamma/\sqrt{M}\>\<\gamma/\sqrt{M}|^{\otimes M}
  e^{-\frac{|\gamma|^2}{2\bar n'}} = \int\frac{M d^2\gamma}{\pi\bar
    n'} |\gamma\>\<\gamma|^{\otimes M} e^{-\frac{M|\gamma|^2}{\bar
      n'}}
\label{finalnumb}
,
\end{split}
\end{equation}
where
\begin{equation}
  2\bar n'+1=\frac1{\nu^2}\left[\mu^2(2\bar n+1)+1\right]=\frac{2M\bar n+2M-N}{N}.
\label{finalnumb2}
\end{equation}
The above state is permutation-invariant and separable, with thermal
local state at each oscillator with average thermal photon
\begin{equation}
  \bar n''=\frac{\bar n'}{M}=\frac{M\bar n+M-N}{MN}\,.
\label{super}
\end{equation}
More generally, for any state $\Xi$, the choice (\ref{bcho}) gives $M$
identical clones whose state can be written as
\begin{eqnarray}
  \rho ' = \int \frac{d^2 \alpha }{\pi}\,e ^{-\frac{|\alpha |^2}{2}(\frac
    1N -\frac 2M +1)}\, 
  \{\Tr [\Xi D^\dag (\alpha  /N)^{\otimes N}] \} \, D(\alpha )
  \;.\label{rhogen}
\end{eqnarray}
Since for any oscillator $c$ one has
\begin{eqnarray}
  \Delta x_c ^2 +\Delta y_c ^2  = \frac 12 + \< c^\dag c \> -|\<c\>|^2,
\label{anym}
\end{eqnarray}
it is easy to verify that the superbroadcasting condition (output
total noise in conjugate quadratures smaller than the input one), is
equivalent to require a smaller number of thermal photons at the
output than at the input, namely
\begin{equation}
  \bar n\geq\frac{M\bar n+M-N}{MN}\,
  \quad\Leftrightarrow \quad \bar n\geq\frac{M-N}{M(N-1)}\,.
\end{equation}
This can be true for any $N>1$, and to any $M\le\infty$, since
\begin{equation}
  \lim_{M\to\infty}\frac{M-N}{M(N-1)}=\frac1{N-1}>0\,.
\end{equation}

\section{Proof of optimality for the channel in Eq. (\ref{optmap})}
\label{optim}
Actually, the solution given in Eq. (\ref{super}) is optimal. To prove
this, in the following we will show that the expectation of the total
number of photons $\Tr[\sum _{l=0}^{M-1} b_l^\dag b_l \map
B(\rho_0^{\otimes N})]$ of the $M$ clones of $\rho$ cannot be smaller
than $M \bar n''$. Since the multisplitter preserves the total number
of photons we have to consider the trace
\begin{eqnarray}
W \doteq \Tr\left [\left(\sum _{l=0}^{M-1} b_l^\dag b_l
\otimes  (U^\dag_a \rho_0^{\tau\otimes N}U_a) \right) S_{a_0b_0}(I_{b_0}\otimes B) S^\dag _{a_0 b_0}\right].\label{vutot}
\end{eqnarray}
We can write $W =W_0 + \sum _{l=1}^{M-1} W_l$, with
\begin{eqnarray}
  &&W_0 \doteq \Tr\left [S^\dag _{a_0 b_0} \left((b_0^\dag b_0\otimes I_{b/b_0})
      \otimes (U_a^\dag\rho_0^{\tau\otimes N}U_a)\right) S_{a_0b_0}
    (I_{b_0}\otimes B) \right],\nonumber\\
  &&W_l \doteq 
  \Tr\left [S^\dag _{a_0 b_0}\left((I_{b/b_l}\otimes b^\dag _l b_l )\otimes 
      (U_a^\dag\rho_0^{\tau\otimes N}U_a)\right) S_{a_0b_0}
     (I_{b_0}\otimes B)
  \right],\label{vuzel}
\end{eqnarray}
for $\ 1\leq l\leq M-1$. Now, since $W_l\geq0$, $W\geq W_0$.
Moreover, using the identity $c^\dag c = -
\partial _{\alpha \alpha ^*}e^{\frac{|\alpha |^2}{2}}D_c( \alpha ) |
_{\alpha =\alpha ^*=0}$, one obtains
\begin{eqnarray}
  &&\Tr_{b_0} \left [ S^\dag _{a_0 b_0} \left(b^\dag _0 b_0
      \otimes \sigma\right) S_{a_0b_0} \right] \nonumber \\& & 
  =- \partial _{\alpha \alpha^* }\int \frac {d^2
    \gamma }{\pi } \Tr _{b_0} [D_{b_0}(\mu \alpha - \nu \gamma ^*)\otimes
  D_{a_0}(\mu \gamma -\nu \alpha ^*)]\, \Tr[D(\gamma)^\dag\sigma]e^{\frac {|\alpha |^2}{2}} |_{\alpha =
    \alpha ^*=0}
  \nonumber \\& & =\left.
    - \frac {1}{\nu ^2}
    \partial _{\alpha \alpha ^*} e^{-\frac {|\alpha |^2}{\nu ^2}}\,e^{\frac {\alpha ^*}{\nu }a_0 ^\dag }\,
    e^{-\frac {\alpha }{\nu }a_0 } \Tr\left[e^{\frac {\mu\alpha ^*}{\nu }a_0 ^\dag }\,
      e^{-\frac {\mu\alpha }{\nu }a_0 }\sigma\right] \right|_{\alpha =\alpha ^*=0} 
  =\frac {a_0^\dag a_0 +\mu ^2\Tr[a^\dag_0a_0\sigma] +1}{\nu ^4},\label{squiquiz}
\end{eqnarray}
then, from Eq.~\eqref{trb} and positivity of $B$, one has
\begin{equation}
\begin{split}
  W_0=&\frac{\Tr [(I_{b/b_0} \otimes a^\dag_0a_0\otimes I_{a/a_0})
    \{I_{b/b_0}\otimes(U_a\rho _0 ^{\otimes N}U_a^\dag)^\tau\}
    B]}{\nu^4}\\
  &\frac{\mu^2\Tr[(I_{b/b_0}\otimes I_{a_0}\otimes \Tr_{a_0}[a^\dag_0
    a_0 (U_a\rho_0^{\otimes N}U_a^\dag)^\tau])B] +\nu^2}{\nu ^4}\\
  &\geq \frac{\mu ^2\bar n +1}{\nu ^2} = \frac NM \bar n +\frac
  {M-N}{N}=M \bar n''.\label{bnd}
\end{split}
\end{equation}
In fact, one can easily check that the choice of $B$ in Eq.
(\ref{bcho}) saturates the bound (\ref{bnd}). Notice that for $\bar
n=0$ one has $N$ coherent states at the input, and $\bar
n''=\frac{M-N}{MN}$, namely one recovers the optimal cloning for
coherent states of Ref. \cite{cerfbrauns}.\par

Also the more general solution given in Eq. (\ref{rhogen}) for
arbitrary input state is optimal, in the sense that it represents the
state of $M$ identical clones with minimal photon number, which is
given by
\begin{eqnarray}
  \Tr [b^\dag b \rho ' ]= \frac {\Tr [
    a^\dag a \rho _0 ]}{N}+\frac 1N -\frac 1M
  \;.
\end{eqnarray}

\section{exact broadcasting}
\label{broad}
In the previous section we proved that it is possible to
superbroadcast thermal states provided that $\bar
n\geq\frac{M-N}{M(N-1)}$. Here we will prove that in this case it is
also possible to broadcast perfectly, namely with local fidelity equal
to 1. Suppose indeed that the Choi-Jamio\l kowski operator is given by
$R_{\map B}=(U_b\otimes U_a)S_{a_0b_0}(I_b\otimes
B')S^\dag_{a_0b_0}(U_b^\dag\otimes U_a^\dag)$, with
\begin{equation}
B'=\nu^2\left(\rho_{\bar m}^{\otimes M-1}\right)_{b/b_0}\otimes |0\>\<0|_{a_0}\otimes I_{a/a_0}.
\end{equation}
Then by using Eq.~\eqref{therm} we replace the first line of
Eq.~\eqref{finalnumb} by
\begin{equation}
\begin{split}
\map B\left(\rho_0^{\otimes N}\right)=&\int\frac{d^2\gamma_0\dots d^2\gamma_{M-1}}{\pi^M} U_b
  [D_{b_0} (\gamma_0)\otimes D(\gamma_1)\otimes\dots\otimes D(\gamma_{M-1})]U_b^\dag\\
& \times e^{-\frac{|\gamma_0|^2}2(2\bar n'+1)-\frac{\sum_{k=1}^{M-1}|\gamma_k|^2}2(2\bar m+1)}=\\
&\int\frac{d^2\tilde\gamma_0\dots d^2\tilde\gamma_{M-1}}{\pi^M}[D_{b_0} (\tilde\gamma_0)\otimes D(\tilde\gamma_1)\otimes\dots\otimes D(\tilde\gamma_{M-1})]\\
& \times e^{-\frac{\left|\frac1{\sqrt M}\sum_{i=0}^{M-1}\tilde\gamma_i\right|^2}{2}(2\bar n'-2\bar m)-\frac{\sum_{k=0}^{M-1}|\tilde\gamma_k|^2}2(2\bar m+1)},
\end{split}
\end{equation}
where $\tilde \gamma_k=\frac1{\sqrt M}\sum_{j=0}^{M-1}e^{\frac{2\pi i
    jk}M}\gamma_j$, and the Jacobian for the change of variables is
clearly 1. Taking the partial trace over all output spaces but one, we
get the following local state on the $k$-th oscillator
\begin{equation}
\rho'=\int\frac{d^2\gamma}{\pi}D(\gamma)e^{-\frac{|\gamma|^2}2(2\bar m+1+\frac{2\bar n'-2\bar m}M)},
\end{equation}
which is a thermal state with thermal photon number
\begin{equation}
\bar n''=\frac{\bar n'+\bar m(M-1)}M=\frac{\bar n}N+\frac1N-\frac1M+\bar m\frac{M-1}M.
\end{equation}
In order to have $\bar n''=\bar n$ it is sufficient to choose a
suitable thermal number $\bar m$, which is the solution of the
following equation
\begin{equation}
\bar n=\frac{\bar n}N+\frac1N-\frac1M+\bar m\frac{M-1}M,
\end{equation}
namely
\begin{equation}
\bar m=\frac{M(N-1)\bar n}{N(M-1)}-\frac{(M-N)}{N(M-1)},
\end{equation}
which is positive iff
\begin{equation}
\bar n\geq\frac{M-N}{M(N-1)}.
\end{equation}
This definitely proves that perfect broadcasting is possible provided
that superbroadcasting is. On the other hand, since superbroadcasting
provides the maximum output purity, it is impossible to have perfect
broadcasting when $\bar n<\frac{M-N}{M(N-1)}$, namely either
superbroadcasting and perfect broadcasting are both possible, or they
are both impossible.

\section{Noise in conjugate quadratures}
We will consider now an alternative derivation, which involves
radiation modes and makes use of a theorem for linear amplifiers. We
are interested in a transformation that provides $M$ (generally
correlated) modes $b_0,b_1,...,b_{M-1}$ from $N$ uncorrelated modes
$a_0,a_1,...,a_{N-1}$, such that the unknown complex amplitude is
preserved and the output has minimal phase-insensitive noise. In
formula, we have input uncorrelated modes
\begin{eqnarray}
&& \< a_ i \>=\alpha \;, \nonumber \\& & 
\Delta x_{a_i} ^2 +\Delta y_{a_i} ^2 = \gamma_i \geq \frac 12, 
\label{}
\end{eqnarray}
for all $i=0,1,..,N-1$, where Heisenberg uncertainty relation is taken
into account. The output modes should satisfy
\begin{eqnarray}
&& \< b_ i \>=\alpha \;, \nonumber \\& & 
\Delta x_{b_i} ^2 +\Delta x_{b_i} ^2 = \Gamma \geq \frac 12
\;, \label{}
\end{eqnarray}
and we look for the minimal $\Gamma $. The minimal $\Gamma $ can be
obtained by applying a fundamental theorem for phase-insensitive
linear amplifiers \cite{caves}: the sum of the uncertainties of
conjugated quadratures of a phase-insensitive amplified oscillator with
(power) gain $G$ is bounded as follows.
\begin{eqnarray}
  \Delta X_B ^2 +\Delta Y_B ^2  \geq G
  (\Delta X_A ^2 +\Delta Y_A ^2  ) + \frac {G-1}{2}
  \;,\label{bou}
\end{eqnarray}
where $A$ and $B$ denotes the input and the amplified mode,
respectively. Our transformation can be seen as a phase-insensitive
amplification from the mode $A=\frac {1}{\sqrt N}\sum _{i=0}^{N-1}
a_i$ to the mode $B=\frac {1}{\sqrt M}\sum _{i=0}^{M-1} b_i$ with gain
$G=\frac MN$, and hence Eq. (\ref{bou}) should hold. Notice that
generally for any mode $c$ one has
\begin{eqnarray}
  \Delta x_c ^2 +\Delta y_c ^2  = \frac 12 + \< c^\dag c \> -|\<c\>|^2
  \;.\label{bou2}
\end{eqnarray}
Hence, the bound can be rewritten as
\begin{eqnarray}
\< B^\dag B \> -|\<B\>|^2 \geq G(\< A^\dag A \> +1 -|\<A\>|^2) -1.
\end{eqnarray}
In the present case, since modes $a_i$ are uncorrelated, one has 
\begin{eqnarray}
&&\<A^\dag A\>= \frac 1N \sum _{i,j=0}^{N-1} \<a^\dag _i a_j\>= 
\frac 1N \left (\sum _{i=0}^{N-1} \<a^\dag _i a_i\>+ 
\sum _{i\neq j} \<a^\dag _i a_j\>\right )\nonumber \\& & 
= (\gamma + |\alpha |^2 -\frac 12)+
(N-1) |\alpha |^2 =\gamma +N |\alpha |^2 -\frac 12,
\end{eqnarray}
where $\gamma=\frac1N\sum_{i=0}^{N-1}\gamma_i$, and so the bound
Eq.~\eqref{bou} is written as
\begin{eqnarray}
\< B^\dag B \> 
\geq G (\gamma +\frac 12)-1+M |\alpha |^2.\label{ass}
\end{eqnarray}
On the other hand, one has 
\begin{eqnarray}
\<B^\dag B\>= \frac 1M \sum _{i,j=0}^{M-1} \<b^\dag _i b_j\>\leq 
\frac 1M \sum _{i,j=0}^{M-1} \sqrt{\<b^\dag _i b_i\> \<b^\dag _j b_j\> 
}= M(\Gamma + |\alpha |^2 -\frac 12).\label{ass2}
\end{eqnarray}
Eqs. (\ref{ass}) and (\ref{ass2}) together give the bound for the
minimal noise $\Gamma$ 
\begin{eqnarray}
\Gamma -\frac 12 \geq  \frac 1N (\gamma -\frac 12) +\frac {1}{N}-
\frac {1}{M}.
\end{eqnarray}
The example in the previous sections corresponds to $\gamma = {\bar
  n}+ \frac 12$ and $\Gamma ={\bar n ''}+\frac 12$. A similar
derivation gives a bound for purification, where $N > M$.  In such a
case $G<1$, and Eq. (\ref{bou}) is replaced with
\begin{eqnarray}
\Delta X_B ^2 +\Delta Y_B ^2  \geq G
(\Delta X_A ^2 +\Delta Y_A ^2  ) + \frac {1-G}{2}
\;,\label{bou2}
\end{eqnarray}
and one obtains the bound
\begin{eqnarray}
\Gamma -\frac 12 \geq  \frac 1N (\gamma -\frac 12) 
\;.\label{pur}
\end{eqnarray}
We would like to stress that the derivation of all bounds in the
present section relies on the theorem of the added noise in {\em
  linear} amplifiers, namely only linear transformations of modes are
considered.  Hence, in principle, these bounds might be violated by
more exotic and nonlinear transformations. Therefore, the derivation
of Eq.  (\ref{bnd}) is stronger, since it has general validity.\par

By a similar derivation, using the bound for phase-conjugated
amplifiers $\Delta X_B^2+\Delta Y_B^2\geq G(\Delta X_A^2+\Delta
Y_A^2)+\frac{G-1}2$, one can obtain the bound for phase-conjugation
broadcasting 
\begin{equation}
\Gamma-\frac12\geq\frac1N\left(\gamma+\frac12\right).
\end{equation}

\section{Experimental implementation}
\begin{figure}[h]
  \epsfig{file=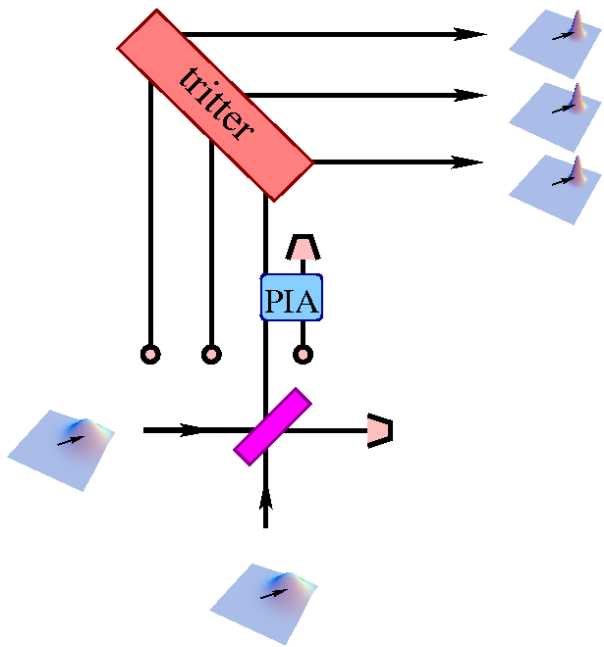,width=8cm}
  \caption{Experimental scheme to achieve optimal superbroadcasting
    from 2 to 3 copies. This setup involves just a beam splitter, a
    phase-insensitive amplifier and a tritter, which in turn can be
    implemented by two suitably balanced beam splitters. The
    phase-insensitive amplifier can be implemented by a beam splitter
    and heterodyne-assisted feed-forward. The output copies carrying
    the same signal as the input ones are locally more pure, the noise
    being shifted to classical correlations between them.\label{sch}}
\end{figure}

The optimal broadcasting can be easily implemented on radiation modes,
by means of an inverse $N$-splitter which concentrates the signal in
one mode and discards the other $N-1$ modes. The mode is then
amplified by a phase-insensitive amplifier with power gain $G=\frac
MN$. Finally, the amplified mode is distributed by mixing it in an
$M$-splitter with $M-1$ vacuum modes. Each mode is then found in the
state of Eq.  (\ref{rhogen}). In the concentration stage the $N$ modes
with amplitude $\<a_i\>=\alpha $ and noise $\Delta x_i^2 +\Delta y_i^2
=\gamma_i$ are reduced to a single mode with amplitude $\sqrt N \alpha
$ and noise $\gamma $. The amplification stage gives a mode with
amplitude $\sqrt M \alpha $ and noise $\gamma ' =\gamma \frac MN +
\frac {M}{2N} -\frac 12$.  Finally, the distribution stage gives $M$
modes, with amplitude $\alpha $ and noise $\Gamma = \frac 1M \left
  (\gamma ' +\frac {M-1}{2}\right )$ each. In Fig. \ref{sch} we sketch
the scheme for $2$ to $3$ superbroadcasting.\par

In Ref. \cite{leuchs} it was shown experimentally that phase
insensitive amplification can be obtained by a setup consisting of a
beam-splitter, a heterodyne detector and a conditional displacement.
In the following we give an algebraic derivation of this result.
Consider a mode in a state $\rho=\int\frac{d^2\gamma}\pi
f(\gamma)D(\gamma)$ coupled to another mode in the vacuum through a
beam-splitter with transmissivity $\tau $. The output is given by the
bipartite state $\sigma$
\begin{equation}
\sigma=\int\frac{d^2\beta d^2\gamma}{\pi^2}e^{-\frac{|\tau \beta-\sqrt{1-\tau ^2}\gamma|^2}2} f(\tau \gamma+\sqrt{1-\tau ^2}\beta)D(\gamma)\otimes D(\beta),
\end{equation}
where we performed the change of variables
$\beta\to\tau\beta+\sqrt{1-\tau^2}\gamma$,
$\gamma\to\tau\gamma-\sqrt{1-\tau^2}\beta$. Now, the reflected mode is
measured by heterodyne detection, and conditionally on the measurement
outcome $\alpha$, a displacement $D(k\alpha)$ is performed on the
transmitted mode, whose state is then given by
\begin{align}
  \rho'= &\int\frac{d^2\alpha d^2\beta d^2\gamma}{\pi^3}e^{-\frac{|\tau\beta-\sqrt{1-\tau^2}\gamma|^2}2} f(\tau\gamma+\sqrt{1-\tau^2}\beta)D(k\alpha)D(\gamma)D(k\alpha)^\dag\<0|D(\alpha)^\dag D(\beta)D(\alpha)|0\>=\nonumber\\
  &\int\frac{d^2\alpha d^2\beta d^2\gamma}{\pi^3}e^{-\frac{|\tau\beta-\sqrt{1-\tau^2}\gamma|^2}2} f(\tau\gamma+\sqrt{1-\tau^2}\beta)e^{\alpha(k\gamma^*-\beta^*)-c.c.}D(\gamma)e^{-\frac{|\beta|^2}2}=\nonumber\\
  &\int\frac{d^2\beta d^2\gamma}{\pi^2}\delta^{(2)}(k\gamma-\beta)e^{-\frac{|\tau\beta-\sqrt{1-\tau^2}\gamma|^2}2} f(\tau\gamma+\sqrt{1-\tau^2}\beta)e^{-\frac{|\beta|^2}2}=\nonumber\\
  &\int\frac{d^2\gamma}{\pi}f(\gamma(\tau+k\sqrt{1-\tau^2}))e^{-\frac{|\gamma|^2}2[k^2+(k\tau-\sqrt{1-\tau^2})^2]}D(\gamma).
\end{align}
On the other hand, the action of a phase-insensitive amplifier on
$\rho$ can be easily calculated and produces the partial output state
\begin{equation}
\rho''=\int\frac{d^2\gamma}{\pi}e^{-\frac{|\gamma|^2\nu^2}2}f(\mu\gamma)D(\gamma).
\end{equation}
The following conditions
\begin{equation}
  \mu=\tau+k\sqrt{1-\tau^2},\quad\nu^2=k^2+(k\tau-\sqrt{1-\tau^2})^2,\quad\mu^2-\nu^2=1,
\end{equation}
which equivalent to
\begin{equation}
  k=\nu,\quad\tau=\frac1\mu,
\end{equation}
imply that $\rho'=\rho''$.Hence, by tuning the beam splitter
transmissivity and the parameter of the conditional displacement $k$,
one can then simulate the amplifier by a linear device assisted by
heterodyne and feed-forward.\par

\section{Conclusion}
\label{concl}

In conclusion, we proved that broadcasting of $M$ copies of a mixed
radiation state starting from $N<M$ copies is possible, even with
lowering the total noise in conjugate quadratures. Since the noise
cannot be removed without violating the quantum data processing
theorem, the price to pay for having higher purity at the output is
that the output copies are correlated. Essentially noise is moved from
local states to their correlations, and our superbroadcasting channel
does this optimally.  We obtained similar results also for
purification (i.e.  $M\leq N$), along with the case of simultaneous
broadcasting and phase-conjugation, with the output copies carrying a
signal which is complex-conjugated of the input one.  Despite the role
that correlations play in this effect, no entanglement is present in
the output (as long as the single input copy has a positive
$P$-function), as it can be seen by the analytical expression of the
output states.  Moreover, a practical and very simple scheme for
experimental achievement of the maps has been shown, involving mainly
passive media and only one parametric amplifier. The superbroadcasting
effect has a relevance form the fundamental point of view, opening new
perspectives in the understanding of correlations and their interplay
with noise, but may be also promising from a practical point of view,
for communication tasks in the presence of noise.

\acknowledgments This work has been supported by Ministero Italiano
dell'Universit\`a e della Ricerca (MIUR) through FIRB (bando 2001) and
PRIN 2005.


\begin{thebibliography}{99}
\bibitem{noclon} W. K. Wootters and W. H.  Zurek, {\em Nature} {\bf 299}, 802 (1982); D.~Dieks,
  Phys. Lett. A, {\bf 92}, 271 (1982); H. P. Yuen, Phys. Lett. A {\bf 113}, 405 (1986); G. C.
  Ghirardi, referee report of N. Herbert, Found.  Phys.  {\bf 12}, 1171 (1982).
\bibitem{buzhill} V. Bu\v zek and M. Hillery, Phys. Rev. A {\bf 54},
  1844 (1996).
\bibitem{gismass} N. Gisin and S. Massar, Phys. Rev. Lett. {\bf 79},
  2153 (1997).
\bibitem{bruss} D. Bruss, D. P. DiVincenzo, A. Ekert, C. Fuchs, C.
  Macchiavello, and J. Smolin, Phys.  Rev. A {\bf 57}, 2368 (1998).
\bibitem{werner} R. F. Werner, Phys. Rev. A {\bf 58}, 1827 (1998); R.
  F. Werner and M. Keyl, J. Math. Phys. {\bf 40}, 3283 (1999).
\bibitem{darlop} G. M. D'Ariano and P. Lo Presti, Phys. Rev. A {\bf
    64}, 042308 (2001).
\bibitem{darmacc} G. M. D'Ariano and C. Macchiavello, Phys. Rev. A
  {\bf 87}, 042306 (2003).
\bibitem{cerf} N. J. Cerf, A. Ipe, and X. Rottenberg, Phys. Rev. Lett.
  {\bf 85}, 1754 (2000); N. J. Cerf and S. Iblisdir, Phys. Rev. Lett.
  {\bf 87}, 247903 (2001).
\bibitem{nobro} H. Barnum, C. M. Caves, C. A. Fuchs, R. Jozsa, and B.
  Schumacher, Phys.  Rev. Lett. {\bf 76}, 2818 (1996).
\bibitem{prl} G. M. D'Ariano, C. Macchiavello, and P. Perinotti, Phys.
  Rev. Lett. {\bf 95}, 060503 (2005).
\bibitem{pra} F. Buscemi, G. M. D'Ariano, C. Macchiavello, and P.
  Perinotti, submitted to Phys. Rev. A.
\bibitem{njp} G. M. D'Ariano, P. Perinotti, and M. F. Sacchi, New J.
  Phys. {\bf 8}, 99 (2006).
\bibitem{epl} G. M. D'Ariano, P. Perinotti, and M. F. Sacchi, accepted
  for publication on Europhys. Lett., eprint number quant-ph/0602037.
\bibitem{ula}U. L. Andersen, R. Filip, J. Fiur\'a \v sek, V. Josse,
  and G. Leuchs, Phys. Rev. A {\bf 72}, 060301 (2005).
\bibitem{bm}S. M. Barnett and P. M. Redmore, {\em Methods in
    Theoretical Quantum Optics}, (Clarendon Press, Oxford, 2002). 
\bibitem{cerfbrauns} S. L. Braunstein, N. J. Cerf, S. Iblisdir, P. van
  Loock, and S. Massar, Phys. Rev. Lett. {\bf 86}, 4938 (2001).
\bibitem{caves} C. M. Caves, Phys. Rev. D {\bf 26}, 1817 (1982).
\bibitem{leuchs} P. K. Lam, T. C. Ralph, E. H. Huntington, and H. A.
  Bachor, Phys. Rev. Lett. {\bf 79}, 1471 (1997).
\end{thebibliography}
\end{document}